# Observation of quantum Hall plateau-plateau transition and scaling behavior of the zeroth Landau level in graphene p-n-p junctions

Cheng-Hua Liu[1,2], Po-Hsiang Wang[2], Tak-Pong Woo[1], Fu-Yu Shih[1,2], Shih-Ching Liou[1,2], Po-Hsun Ho[3], Chun-Wei Chen[3], Chi-Te Liang[1], and Wei-Hua Wang[2*]

[1]Department of Physics, National Taiwan University, Taipei 106, Taiwan

[2]Institute of Atomic and Molecular Sciences, Academia Sinica, Taipei 106, Taiwan

[3]Department of Materials Science and Engineering, National Taiwan University, Taipei 106, Taiwan

[*]Corresponding Author. (W.-H. Wang) Tel: +886-2-2366-8208, Fax: +886-2-2362-0200; E-mail: wwang@sinica.edu.tw

**Abstract**

We report distinctive magnetotransport properties of a graphene p-n-p junction prepared by controlled diffusion of metallic contacts. In most cases, materials deposited on a graphene surface introduce substantial carrier scattering, which greatly reduces the high mobility of intrinsic graphene. However, we show that an oxide layer only weakly perturbs the carrier transport, which enables fabrication of a high-quality graphene p-n-p junction through a one-step and resist-free method. The measured conductance-gate voltage ($G - V_G$) curves can be well described by a metal contact model, which confirms the charge density depinning due to the oxide layer. The graphene p-n-p junction samples exhibit pronounced quantum Hall effect, a well-defined transition point of the zeroth Landau level (LL), and scaling behavior. The scaling exponent obtained from the evolution of the zeroth LL width as a function of temperature exhibits a relatively low value of $\kappa = 0.21 \pm 0.01$. Moreover, we calculate the energy level for the LLs based on the distribution of plateau-plateau transition points, further validating the assignment of the LL index of the QH plateau-plateau transition.

Electrical transport studies of graphene heterostructures[1] have revealed the quantum Hall effect (QHE),[2] quantum interference behaviors,[3,4] Klein tunneling,[5,6] and the split closed-loop resonator,[7] hence convincingly demonstrating the advantages of constructing in-plane heterostructures of graphene. Specifically, the relativistic quantization of graphene's electronic spectrum results in distinct characteristics, including the presence of a Landau level (LL) at zero energy and chiral QHE.[8-12] In the quantum Hall (QH) regime, the plateau-plateau transition and the scaling behavior have been studied using Hall bar[13-18] and Corbino geometry,[19] providing important information about the carrier localization in graphene. A scaling exponent for typical Anderson-type transition $\kappa \cong 0.42$ has been reported in graphene,[13,14] while some studies showed reduced value of $\kappa$.[15-19] In a graphene Hall bar, the slope of the Hall conductivity at the transition region, $d\sigma_{xy}/d\nu$, exhibits scaling behavior with $\kappa = 0.41$ for the first and second LLs, while that of the zeroth LL is temperature ($T$) independent.[13] In a Corbino geometry, FWHM $\Delta\nu$ of the zeroth LL is $T$ dependent and shows scaling behavior with $\kappa = 0.16$, which is attributed to an inhomogeneous charge carrier distribution.[19] However, until now, a well-defined QH plateau-plateau transition point of the zeroth LL of graphene has not been directly observed, obscuring a detailed understanding of the scaling behavior of this unique LL. Moreover, the variation of the $T$ exponent in previous reports suggests the need for further investigation of the scaling behavior of the zeroth LL using different methods for the device structures.

In this report, we fabricate a high-quality graphene p-n-p junction, achieved via controlled diffusion of metallic contacts, to explore the QH plateau-plateau transition and scaling behavior of graphene. Interestingly, we observe a well-defined transition point corresponding to the zeroth LL, revealing the scaling behavior and a reduced $T$ exponent. There are additional advantages of utilizing a graphene p-n-p junction to explore the transition region of the QHE. First, the presence of the transition between an integer QH plateau and a QH plateau with a fractional value enables direct access to the transition of the zeroth LL. Second, in a graphene p-n-p geometry, the intrinsic graphene is adjoined by the doped graphene from both sides. The doped graphene regions can be viewed as an ideal contact, facilitating the investigation of the transition region of the intrinsic graphene. Moreover, we derive the value of energy level for the observed LLs, which agrees with the theoretical values, further validating the assignment of the LL index of the QH plateau-plateau transition.

A detailed device fabrication process can be found in supporting information S1. Briefly, we exfoliated graphene onto the $SiO_2$/Si substrates modified by the organic molecule octadecyltrichlorosilane (OTS), which can greatly reduce charged-impurity scattering and provide an ultra-smooth substrate surface.[20] We then employed resist-free fabrication with a shadow mask to reduce possible polymer residue. A crucial step in this process was to control the metal diffusion of electrodes by deliberately increasing the gap between the shadow mask and graphene samples.[21] Figure 1a shows a schematic of the graphene p-n-p junction device with a pronounced diffusion of the metallic contact. With a gap of 130±5 µm, we deposited 5-nm-thick Ti and 50-nm-thick Au as contact electrodes, which resulted in a large lateral diffusion of approximately 4 µm, as evidenced by an AFM image of sample A (Figure 1b, blue gradient area). Before the transport/magnetotransport measurement, the samples were annealed at 383 K for 3 hours in a low vacuum (Helium atmosphere) to remove adsorbates.[20]

We note the critical role of the Ti layer in determining the transport behavior by comparing the two-terminal conductance vs. gate voltage ($G - V_G$) curves of a graphene p-n-p junction (sample A) and a control sample (sample B), as shown in Figure 1c and 1d, respectively. For sample A, a large field effect is observed for $V_G < 7$ V, showing a typical graphene characteristic, in which the $G - V_G$ curve can be well described by the self-consistent Boltzmann equation $\rho = (ne\mu_c + \sigma_0)^{-1} + \rho_s$, where $\mu_c$, $\rho_s$, and $\sigma_0$ are density-independent mobility, the resistivity due to short-range scattering, and residual conductance at the Dirac point, respectively. The fitting yields that $\mu_c$ and $\rho_s$ of sample A are 3000 cm$^2$/Vs and 368 Ω, respectively. However, the $G - V_G$ curve for $V_G > 7$ V shows an additional conductance minimum at $V_G = 32$ V, suggesting a doping effect of graphene in the diffused electrode region (supporting information S2).[22,23] The double conductance minimum was observed in other graphene p-n-p junction devices, as shown in supporting information S3. To examine the role of the Ti layer, we have fabricated sample B such that the contact metal is made only with Au instead of the Ti/Au bilayer. Different from sample A, sample B exhibits a typical $G - V_G$ curve in the regime $V_G > V_{CNP}$, as shown in Figure 1d. We therefore infer that the double dip feature observed in sample A is related to the Ti adhesion layer.

It is well understood that the charge density of graphene under metallic contact is pinned due to Fermi level pinning,[24,25] leading to typical transfer characteristics.

Nevertheless, an additional kink is observed in the $G - V_G$ curve, suggesting that the depinning of charge density occurs.[26,27] We performed a theoretical simulation of the $G - V_G$ curves based on a simple metal contact model[26] (supporting information S4). By assuming charge-density depinning and employing the potential profile $V(x)$ depicted in the inset of Figure 1e, we calculated transfer characteristics that reasonably fit the measured data, as shown in Figure 1e. The agreement between the modeled and the measured $G - V_G$ curve suggests that the interfacial metal is oxidized.[27] In the calculation, we applied the device geometry with $L_0/W = 1.06$ and assumed $L_1 = 0.35 L_0$ ($L_0 = 12$ µm). We note that the resulting $L_1 = 4.2$ µm is comparable to the diffusion length of 4 µm obtained from the AFM images (Figure 1b), indicating that the interfacial oxidation occurs approximately in the diffused area. The depinning enables large-area modification of the carrier density,[26,27] which causes a doped graphene region.[28] We therefore conclude that charge-density depinning and the doping effect occur in the diffusion region, leading to a graphene p-n-p structure.

We present the magnetotransport properties of the fabricated graphene p-n-p junctions. Figure 2a compares the $G - V_G$ curves of sample A at $B = 0$ T and $B = 9$ T. The graphene sample exhibits two conductance minima at $B = 0$ T, corresponding to the pristine and doped graphene regions mentioned above. Therefore, sample A is in the unipolar regime for $V_G < 7$ V and $V_G > 32$ V and in the bipolar regime for $7$ V $< V_G <$ $32$ V. At $B = 9$ T, sample A manifests a pronounced QHE, revealing QH plateaus for $\nu = 2, 6, 10$ in the unipolar regime and a QH plateau at $\nu = 2/3$ in the bipolar regime. In a graphene p-n-p junction, the direction of the chiral edge states under a magnetic field is determined by the type of carriers. In the bipolar regime, these chiral states circulate with opposite direction in p-type and n-type regions, as shown in the schematic of Figure 2b. When the edge states form a compressible channel at the p/n interface with full mixing equilibrium,[2] the conductance can be expressed as $G = v_{pnp} e^2/h$, with

$$v_{pnp} = \frac{|v_1||v_2|}{|v_1|+ 2|v_2|} (v_1 v_2 < 0, v_1, v_2 = \pm 2, \pm 6, \pm 10 ...)$$

The mixture of $v_1 = 2$ of the intrinsic graphene region and $v_2 = -2$ of the doped region results in the observed $v = 2/3$ of the first mixing filling factor. We note that the carriers can propagate through the graphene p-n interface with suppressed backscattering known as Klein tunneling,[3,5] which may facilitate the observed full-mixing of the edge states. In the unipolar regime, the filling factor is given by $G = min(|v_1|, |v_2|) \times e^2/h$, which can account for the integer QH plateau of $v = 2, 6, 10$. The edge states circulate in the same direction in all three regions and only the edge states that permeate the whole channel contribute to the measured $G$. The edge states that propagate across the interface is nondissipative due to the suppressed backscattering.[29] It is noted that the observation of the fractional-valued QH plateaus is irrelevant to the factional QH effect due to the formation of complex composite quasiparticles [30,31], but is originated from the merging of the edge state at the p-n interface.

In supporting information S3, we show two other graphene p-n-p junction devices (sample C and D), which exhibit a comparable QHE, indicating the validity of the fabrication method. We note that despite the presence of charged impurity scattering introduced by the oxide layer, the high mobility and the QH regime can still be attained in our graphene samples. To evaluate the disorder in our graphene p-n-p junction devices, we calculate the Ioffe-Regel parameter $(k_F\lambda)^{-1}$, where $\lambda$ is the transport mean free path, yielding $(k_F\lambda)^{-1} = 0.5, 0.3, 0.4$ for sample A, C, and D, respectively. The obtained $(k_F\lambda)^{-1} < 1$ indicates that the graphene p-n-p junction samples have reasonably low disorder. Furthermore, Figure 2d shows the derivative of $dG/dV_G$ as a function of $V_G$ for $2\ \text{K} < T < 100\ \text{K}$. We note that the minimum for the zeroth LL exhibits split peaks, while the minima for other LLs only show a single peak. The presence of the split peaks of the zeroth LL may be attributed to sublattice symmetry breaking in the samples with the small disorder strength,[32] which is consistent with the aforementioned low disorder of the sample.

A schematic diagram of the energy level of the LLs in the graphene p-n-p junction is shown in Figure 2e. Because of the extra carriers induced by the p-type doping, the energy of the LLs in the doped region is higher than that of the same LL index in the intrinsic region. We note that the device does not show a transition when the Fermi level crosses the LLs of the doped graphene because the chiral edge states in those regions reflect back to the same electrodes (see schematics of Figure 2b and 2c) and do not contribute to the two-terminal resistance. Therefore, the transition of the QHE is only manifested by the LLs of the intrinsic graphene. The doped graphene regions thus act as a contact in this sense, offering a unique means to probe the LLs of the intrinsic graphene.

We further discuss the robustness of the QH state against the thermal energy. Figure 3a shows the $G - V_G$ curves for $2\ \text{K} < T < 100\ \text{K}$. At $T = 2\ \text{K}$, the sample exhibits the pronounced QH plateaus at $\nu = 2, 6, 10$ and $\nu = 2/3$ for the unipolar and bipolar regime, respectively. The QH plateau at $\nu = 2$ is particularly robust, persisting up to $T = 100\ \text{K}$.

Conversely, the QH plateaus at ν = 6, 10 and ν = 2/3 are further subjected to the influence of the thermal energy. The energy quantization of graphene in a magnetic field can be written as $E_N = v_F\sqrt{|2e\hbar BN|}$, where $v_F$ is the Fermi velocity and $N$ is the LL index. The energy gap of the QH plateau at ν = 2 can be estimated as $\Delta E = E_1 - E_0 \approx 1000$ K at $B = 9$ T. Because the cyclotron gap, $\hbar\omega_c$, is larger than the thermal energy, $k_B T$, by a factor of 10,[33] the persistence of the QH plateau at ν = 2 for $T < 100$ K, which is comparable to the observed critical $T$, is implied. Alternatively, the energy gap of the fractional-valued QH plateau at ν = 2/3 is smaller than that of the QH plateau at ν = 2 (see Figure 2e), causing a weaker persistence of the QHE against thermal excitation.

We now focus on the scaling behavior of the QH plateau-plateau transitions in our graphene p-n-p junction devices. In the QH regime, both the localized and extended states are critical to the development of the QHE, and the width of the QH plateaus depend on the ratio of localized to extended states.[34] At the QH plateau-plateau transition, the presence of the delocalized states results in nonzero $\sigma_{xx}$; $\sigma_{xy}$ becomes non-quantized because these states condense into a fluid state, leading to a transition region of nonzero width between quantized values.[35] The delocalization can be manifested by the scaling behavior of the magnetoresistance (MR) as a function of $T$ in the transition region,[36] which can be inferred as follows. In the center of a LL, the localization length ξ of electronic states diverges as $\xi \propto |\nu - \nu_c|^{-\gamma}$, where $\nu_c$ is the LL center and ν is its localization edge.[37] It can then be derived that the maximum slope of $d\sigma_{xy}/d\nu$ diverges as $T^{-\kappa}$ in the transition region, with the exponent $\kappa = p/2\gamma$, where γ is the localization length exponent and $p$ is the inelastic scattering exponent.[36,37] Hence, the deviation of the field from its critical value $(\nu - \nu_c)$ rescales by the factor $T^{\kappa}$ as $T$ decreases; it follows that the transition region becomes smaller as $T$ decreases.

From Figure 3a, we can extract $(dG/d\nu)_{max}$, the maximum of the slope for each $T$, in which the filling factor $\nu$ is calculated using the relation $\nu = nh/eB$. We can then plot the $T$ dependence of $(dG/d\nu)_{max}$ for the first and second LLs corresponding to the intrinsic region of the p-n-p junction, as shown in Figure 3b. The value of $(dG/d\nu)_{max}$ is extracted up to the $T$ value at which the QH plateaus are about to disappear. Sample A exhibits a scaling behavior, and $(dG/d\nu)_{max}$ varies linearly with $T$ for 30 K $< T <$ 70 K, yielding $\kappa = 0.36 \pm 0.01$ and $\kappa = 0.35 \pm 0.01$ for the first and second LLs, respectively. In this study, we measured the two-terminal $G$ composed of both $\sigma_{xx}$ and $\sigma_{xy}$ components, which may complicate the analysis of the MR. However, we note that the discussion of the scaling behavior and the extraction of the slope $dG/d\nu$ are restricted in the plateau-plateau transition of the intrinsic region (see Figure 2e), while both $\sigma_{xx}$ and $\sigma_{xy}$ corresponding to the doped region are within the QH plateau regime and do not affect the value of $G$. Moreover, although $\sigma_{xx}$ is not zero at the transition of the intrinsic region, it is around the maximum value where $(dG/d\nu)_{max}$ occurs and contribute insignificantly to $dG/d\nu$. Therefore, the measured alteration of the slope $dG/d\nu$ is dominated by the $T$ dependence of $\sigma_{xy}$, which validates the analysis of the scaling behavior of the QH plateau-plateau transition by using two-terminal geometry.

Here, we assume that $p = 2$, which is generally accepted for a 2D system dominated by short-range scattering. This is conceivable because the carrier transport is dominated by short-range scattering for TiO$_2$ on graphene.[38] From the relation $\kappa = p/2\gamma$, we then obtain $\gamma = 2.7$, which is in reasonable agreement with the theoretical calculation $\gamma = 2.35$,[34] indicating that localization of the higher order LLs in our graphene p-n-p samples is governed by a scaling behavior similar to that in conventional 2D systems. At lower $T$, the $T$ dependence of $(dG/d\nu)_{max}$ becomes smaller because the localization length approaches the intrinsic scattering length, which is $T$ independent.[19] We define $T_C$ as the $T$ where the data start to deviate from the scaling behavior. It is found that $T_C$ for the first and second LLs are comparable (approximately 30 K), suggesting that $T_C$ is associated with the dimension of the sample but not the LL index.

Notably, we observe the scaling behavior of the QH plateau-plateau transition between integer and fractional-valued QH plateaus of graphene. Figure 2c shows the $G - V_G$ curves at the transition region between the integer QH plateau $\nu = 2$ and the fractional-valued QH plateau $\nu = 2/3$ for $T$ ranging from 2 K to 20 K. According to the energy levels of the LLs shown in Figure 2e, this transition occurs at the zeroth LL of the intrinsic graphene region. We obtain $\kappa = 0.21 \pm 0.01$ for the zeroth LL, which is smaller than those of the first and second LLs. A reduced value of $\kappa$ for the zeroth LL has been reported in the graphene Corbino device,[19] which has been attributed to the dominance of electron-hole puddles, as evidenced by an inhomogeneous charge carrier distribution.[13,19] Because the transition of the zeroth LL coincides with the charge neutrality point, the reduced $\kappa$ observed in the graphene p-n-p junction may be related to the presence of the electron-hole puddles.[39]

Finally, we estimate the energy level ($E_N$) for the LL index to validate the assignment of the LLs of the QH plateau-plateau transition. Figure 4a shows the MR curves of sample A for different $V_G$ in the unipolar regime. At $-0.3 \text{ T} < B < 0.3 \text{ T}$, we observe negative MR, which can be attributed to weak localization (WL).[40,41] Beyond the WL regime, sample A exhibits positive MR, which can be explained by the classical Hall effect. Moreover, we observe oscillating peaks, which are coupled by the Shubnikov-de Haas (SdH) oscillations in $\sigma_{xx}$ for $-50 \text{ V} < V_G < -10 \text{ V}$. For $V_G = 0 \text{ V}$, sample A enters the QH regime for $B > 3 \text{ T}$ and manifests a pronounced QH plateau at $\nu = 2$, which is consistent with the aforementioned QHE. We then construct the Landau fan diagram by identifying the value of the $1/B_N$ field corresponding to the $N$-th maxima and minima of the SdH oscillations and by plotting against the Landau index $N$ at different $V_G$, as shown in Figure 4b. The $N$ versus $1/B_N$ data at different $V_G$ can be fitted linearly, and these lines extrapolate and converge at $-0.5$ of the y-axis, indicating a nonzero Berry's phase of monolayer graphene.[10] From the slope of the linear fitting, we can obtain the SdH oscillation period $1/B_F(V_G)$, yielding the $V_G$ dependence of carrier density as $n = 4e/hB_F(V_G)$, which is shown in Figure 4c. We found that $n$ varies linearly with $V_G$, and the capacitance of the graphene device on 300-nm-thick $SiO_2$ is calculated as $\alpha = n(V_G)/V_G = 4.77 \times 10^{10} \text{ cm}^{-2}\text{V}^{-1}$.

From the Landau fan diagram, the relationship between $E_F$ and $n(V_G)$ can be derived using $B_F = E_F^2/2ev_F^2\hbar$ [42-44], where $v_F = 10^6$ m/s, as shown in the inset of Figure 4d. We note that $E_F$ varies linearly with the square root of $n$, which is consistent with the behavior of relativistic Dirac particles described by $E_F = \hbar v_F k_F = \hbar v_F \sqrt{n\pi}$.[45,46] We then identify $V_G(\nu)$ for the LL index based on the corresponding derivative minima from $dG/dV_G$ versus $V_G$ curves (Figure 2d). By converting $V_G(\nu)$ to $n(\nu)$ and using the $E_F - n$ relation, we obtain the $E_N$ corresponding to the LL index, as shown in Figure 4d. The estimated $E_N$ versus the LL index agrees well with theoretically derived values based on $E_N = v_F \sqrt{2e\hbar BN}$ with B = 9 T, indicating the validity of the assignment of the LLs of the QH plateau-plateau transition.

In summary, we demonstrated a unique method of fabricating a graphene p-n-p junction by controlling the lateral diffusion of the metallic contacts. The measured $G - V_G$ curves can be thoroughly described by the metal contact model, confirming the charge density depinning and the presence of interfacial oxidation. The graphene p-n-p junction devices showed pronounced QHE, a well-defined transition point of the zeroth LL, and the scaling behavior. We also estimated $E_N$ for the LL index which is consistent with the theoretically derived values. The demonstration of a high-quality graphene p-n-p junction with controlled diffusion of the contacts provides an alternative fabrication method for future graphene-based electronics.

**ACKNOWLEDGMENT**

This work was supported by the Ministry of Science and Technology of Taiwan under contract numbers MOST 103-2112-M-001-020-MY3 and MOST 104-2622-8-002-003.

**Figure captions**

**Figure 1. The structure and the transport characteristics of the graphene p-n-p junctions.** (a) A schematic of the structure of the graphene device with diffused electrode edges. (b) An AFM image showing the edge diffusion of sample A. Comparison of the $G - V_G$ curves between (c) sample A and (d) sample B with Au electrode. (e) Measured $G - V_G$ curve for sample A at $T = 2$ K and simulation based on the metal contact model. Inset: depinning potential assumed in the simulation.

**Figure 2. The magnetotransport of the graphene p-n-p junctions.** (a) The $G - V_G$ curves of sample A at $B = 0$ T (black) and $B = 9$ T (red). (b, c) Edge state circulation of the graphene p-n junctions in the QH regime. (b) Edge currents in p- and n-regions, which circulate in opposite directions and mix at the p-n interface in the bipolar regime. (c) Edge currents in the unipolar regime, which circulate in the same direction. (d) The differential conductance as a function of $V_G$, indicating the plateau-plateau transition points. (e) A schematic of the energy distribution of the LLs of the graphene p-n-p junction.

**Figure 3. The scaling behavior of the QH plateau-plateau transitions.** (a) The $G - V_G$ curves of sample A from $T = 2$ K to $T = 100$ K. (b) $(dG/d\nu)_{max}$ as a function of $T$ and the fits, which yield the scaling exponent for the first and second LLs. (c) The $G - V_G$ curves of sample A from $T = 2$ K to $T = 20$ K, exhibiting the transition region between the integer QH plateau $\nu = 2$ and the fractional-valued QH plateau $\nu = 2/3$. (d) $(dG/d\nu)_{max}$ as a function of $T$ and the fitting of the scaling exponent for the zeroth LL.

**Figure 4. The MR and the distribution of the energy levels of the LLs.** (a) The MR, showing SdH oscillations in the unipolar regime. (b) A fan diagram showing the LL index as a function of $1/B_N$ at different $V_G$. (c) The carrier concentration $n = 4e/hB_F(V_G)$, which is linearly dependent on $V_G$, from which the carrier density capacitance α is extracted. (d) The energy of the $N^{th}$ LL ($E_N$) as a function of the LL index. Inset: Fermi energy as a function of carrier density.

**Figure 1.**

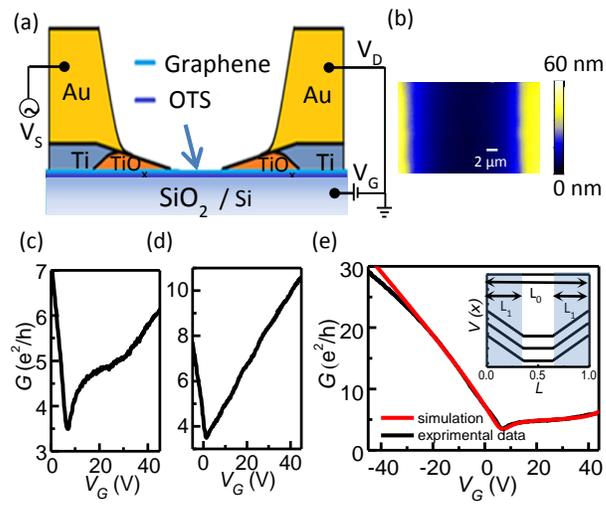

**Figure 2.**

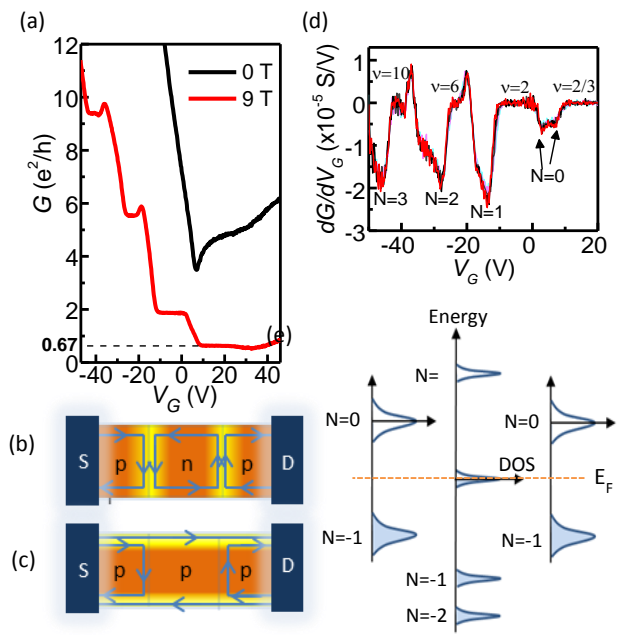

**Figure 3.**

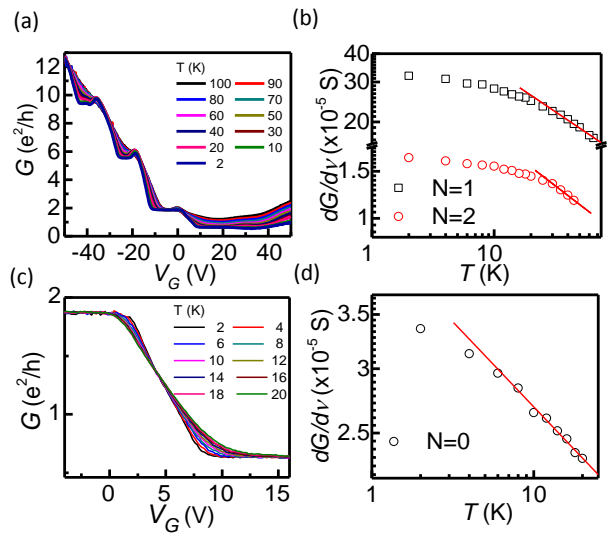

**Figure 4.**

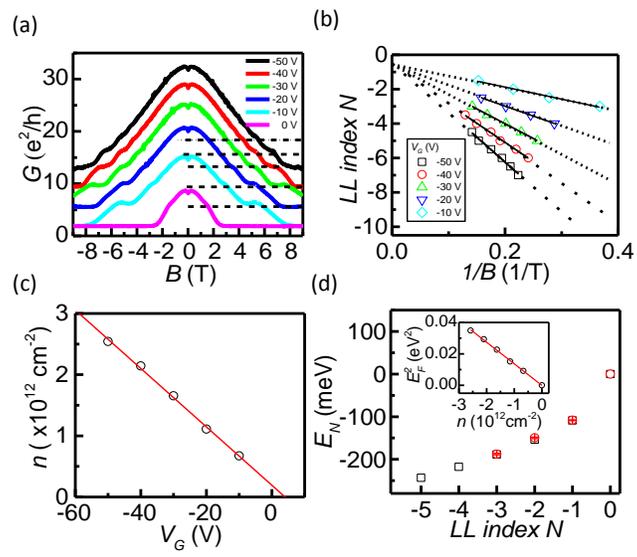

# Supplemental Material for

**Observation of quantum Hall plateau-plateau transition and scaling behavior of the zeroth Landau level in graphene p-n-p junctions**

Cheng-Hua Liu[1,2], Po-Hsiang, Wang[2], Tak-Pong Woo[1], Fu-Yu Shih[1,2], Po-Hsun Ho[3], Chi-Te Liang[1], Chun-Wei Chen[3], and Wei-Hua Wang[2*]

[1]Department of Physics, National Taiwan University, Taipei 106, Taiwan

[2]Institute of Atomic and Molecular Sciences, Academia Sinica, Taiwan

[3]Department of Materials Science and Engineering, National Taiwan University, Taipei 106, Taiwan

[*]Corresponding Author. (W.-H. Wang) Tel: +886-2-2366-8208, Fax: +886-2-2362-0200; E-mail: wwang@sinica.edu.tw

## S1: Device fabrication and electrical measurement

We exfoliated monolayer graphene onto $SiO_2$/Si substrates modified by a self-assembled monolayer of organic molecules (octadecyltrichlorosilane) [1]. Prior to the metal deposition, we aligned graphene flakes with respect to a shadow mask (a TEM grid) by using a x-y-z micro-manipulator under an optical microscope (OM). We then deposited Ti/Au (5 nm/50 nm) onto the graphene sample as contact electrodes, and the extent of the edge diffusion of the electrodes was fine controlled by a proper distance between the shadow mask and the graphene samples [2]. The graphene p-n-p devices were annealed at 110 °C for 3 hours in a low vacuum (Helium atmosphere) to remove the adsorbates on the surface of graphene, then cooled down to $T = 2$ K. We measured the devices by using a standard lock-in technique, with 100 µV of source-drain voltage and a modulation frequency of 17 Hz. The current signal was amplified by using a current pre-amplifier (Stanford Research Systems, SR570) and then measured by using a lock-in amplifier (Stanford Research Systems, SR830). $V_G$ was applied by using a DC source (Keithley 2400).

## S2: Transport/magnetotransport properties of the control samples

To understand the critical role played by the Ti layer, we fabricated control samples with an Au electrode and deposited Ti (1 nm) onto the graphene channel by using a polymethyl methacrylate (PMMA) stencil mask [6]. Figure S2(a) compares the $G - V_G$ curve of a control sample before and after Ti deposition. The CNP is greatly shifted toward the positive voltage of ~64 V, indicating strong p-type doping. This p-type doping is consistent with the formation of the graphene p-n-p junctions through controlled diffusion as well as with oxidation that causes the charge-density depinning discussed in the main text.

Figure S2(a) shows that the carrier mobility is suppressed for both electron and hole branches after Ti deposition, which can be attributed to the increased charged impurity scattering. Figure S2(b) shows the $G - V_G$ curves of the control sample before Ti deposition at $B = 9$ T, where the conventional integer QH effect is observed. After Ti deposition, the mobility dropped to ~1000 $cm^2/\text{V} \cdot \text{s}$. Nevertheless, the control sample can still attain to a QH regime with this Ti deposition because the QH plateau of $\nu = 2$ is observable at $B = 9$ T, as shown in Figure S2(c), suggesting that a graphene p-n-p junction is feasible with titanium oxide as a doping source.

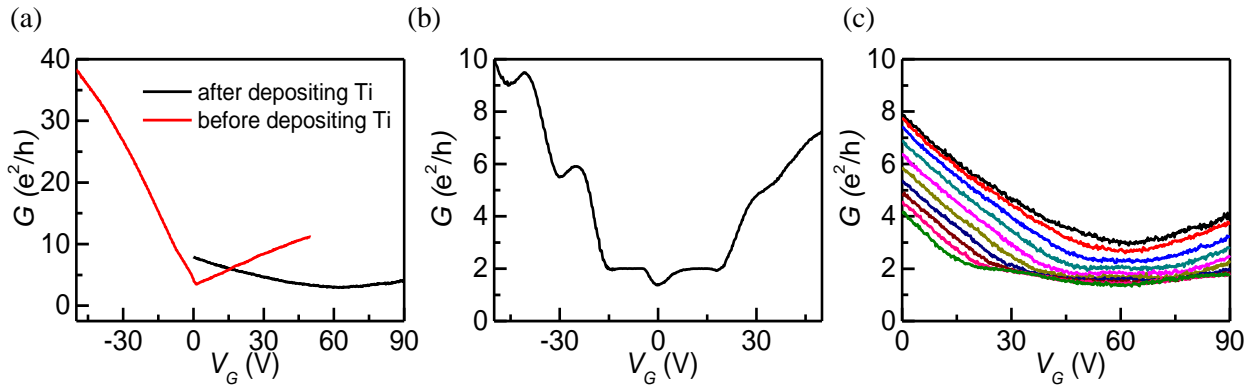

**Figure S2.** (a) $G - V_G$ curves before (red curve) and after (black curve) deposition of Ti onto the graphene channel. (b) The control sample before Ti deposition exhibits QH plateaus of $\nu = 2, 6$ for both electron and hole branches at $B = 9$ T. (c) The control sample after Ti deposition approaches a QH plateau of $\nu = 2$ as the magnetic field increases from $B = 0$ T to $B = 9$ T.

## S3: Characteristics of graphene p-n-p junction devices

We demonstrated the characteristics of three graphene p-n-p devices, including sample A (discussed in the manuscript) and samples C and D (shown here). The extent of the electrode diffusion of samples C and D is comparable to that of sample A. The charged neutrality points (CNP) of samples A, C, and D correspond to 7, 2, and 0 $V$, respectively, as shown in Figure 2 (a) of the main text and Figure S3 (a) and (b). The CNP manifests in the vicinity of zero $V_G$, which is attributed to the low density of charged impurities on the OTS-modified substrates. The field effect mobility in the unipolar regions of samples A, C, and D are 3000, 5308, and 3774 cm$^2$/V·s, respectively. It is expected that the graphene samples attain to the quantum Hall (QH) regime at B = 9 T, based on the condition $\mu B > 1$ [3].

Both samples C and D exhibit QH plateaus of $\nu = 2, 6$ in the unipolar regime and $\nu = 2/3$ in the bipolar regime at B = 9 T, as shown in Figure S3 (a) and (b), suggesting the consistency of the fabrication method. Figure S2 (c) and (d) show the derivative $dG/dV_G$ for samples C and D as a function of $V_G$ for $T = 2$ K. It is found that the minimum for the zeroth LL exhibits split peaks, while the minima for other LL only show a single peak, indicating the qualitatively same behavior for samples C and D. For sample C, the distance between the transition points for the zeroth and first LL is 18.5 V, which is equivalent to 110 meV based on the gate capacitance of $4.77 \times 10^{10}$ cm$^{-2}$V$^{-1}$ calculated in the main text. This energy difference is consistent with the theoretical value of 109 meV derived from $E_N = v_F\sqrt{|2e\hbar BN|}$. Moreover, we estimate the resistivity corresponding to short-range scattering based on the self-consistent Boltzmann equation $\rho = (ne\mu_c + \sigma_0)^{-1} + \rho_s$, where $\mu_c$, $\rho_s$, and $\sigma_0$ are density-independent mobility, the resistivity due to short-range scattering, and residual conductivity at the Dirac point, respectively. $\rho_s$ for samples A, C, and D is extracted as 368 Ω, 224 Ω, and 389 Ω, respectively.

(a)

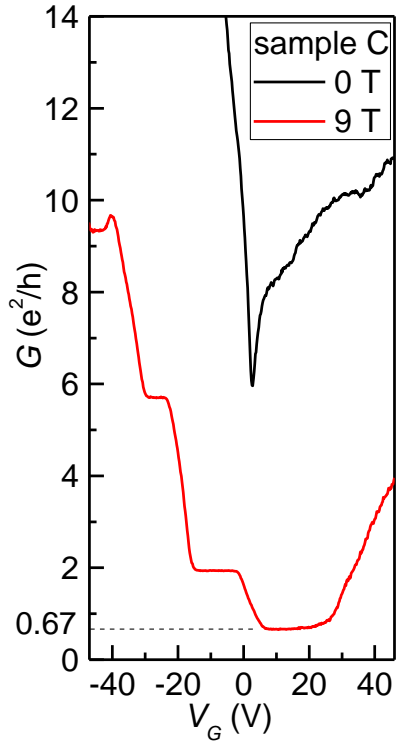

(b)

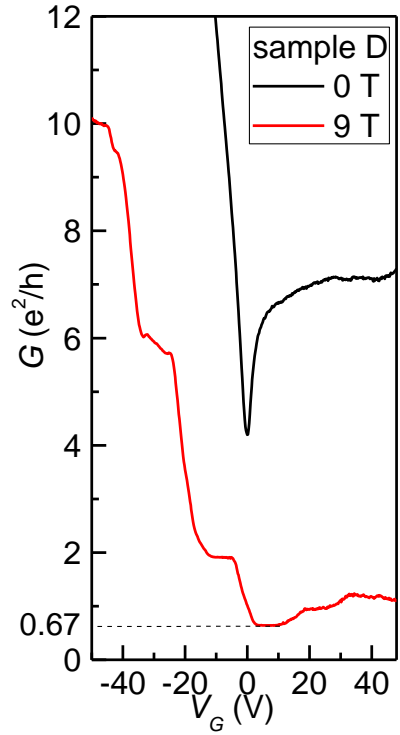

(c)

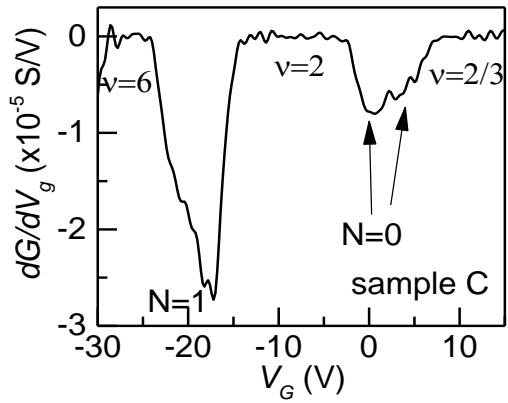

(d)

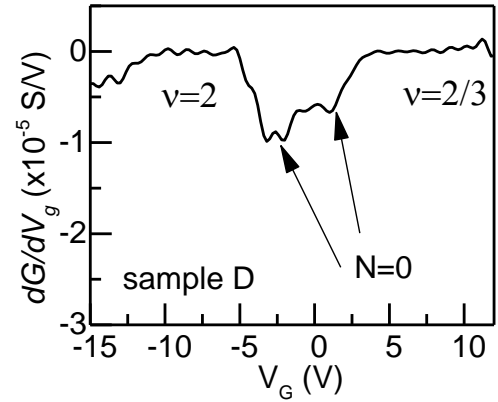

**Figure S3.** (a) (b) The transport and magnetotransport for samples C and D. Both samples exhibit QH plateaus of $\nu = 2, 6$ in the unipolar regime and $\nu = 2/3$ in the bipolar regime. (c) (d) The differential conductivity $dG/dV_G$ as a function of $V_G$ for samples C and D.

## S4: Theoretical simulation based on metal contact model

We confirm the structure of a diffused electrode via theoretical simulation based on a simple metal contact model [5]. The overall channel resistance, $R$, can be written as follows: $R = \frac{1}{W}\int_0^{L_0} \frac{1}{\sigma(x)} dx$, where $L_0$, W and $\sigma(x)$ are the channel length, sample width and local conductivity, respectively. The ratio $L_0/W = 1.06$ is extracted from the sample geometry. The local conductivity can be formulated as $\sigma(x) = \sqrt{\{\mu C_0[V_G - V_D(x)]\}^2 + \sigma_{min}^2}$, where $C_0$ is the gate capacitance per unit area and $\mu$ is the carrier mobility. Therefore, $R$ can be re-written as $R = \frac{1}{W}\int_0^{L_0}(\{\mu C_0[V_G - V_D(x)]\}^2 + \sigma_{min}^2)^{-1/2} dx$. We assume charge-density depinning on the edges and that the potential of the contact regions varies monotonically within the range of $L_1 = 0.35L_0$ on both sides, as shown in the inset of Figure 1(e) of the main text. The simulation yields a reasonable fitting to the experimental data within the range of $-45\,V < V_G < 45\,V$, as shown in Figure 1(e) of the main text. Moreover, the doping region is $L_1 = 0.33L_0$, which agrees well with the diffusion length of 4 μm based on the AFM characterization. In contrast, if the potential is pinned on both side edges [Figure S4(a)], the simulation cannot fit our experimental data, as shown in Figure S4(b). These simulation results suggest that the oxidation occurred in the diffusion area, as discussed in the main text.

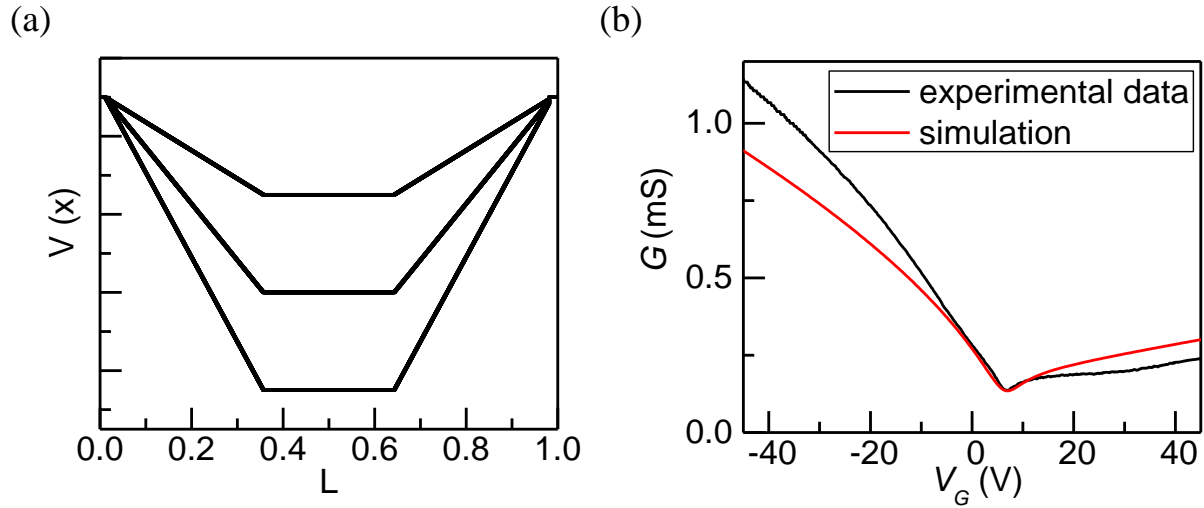

**Figure S4**: (a) The potential profile, assuming charge-density pinning. (b) $G - V_G$ curve and simulation, assuming the pinning potential with the same parameter $L_1 = 0.33L_0$.